\documentclass[a4paper, 12pt]{article}
\usepackage[T1]{fontenc}
\usepackage{lmodern}
\usepackage[latin1]{inputenc}
\usepackage{amsfonts, latexsym, amsmath, amssymb} 
\usepackage[top=2cm, bottom=3cm, left=2cm, right=2cm]{geometry}
\usepackage{graphicx}
\usepackage{bm}
\usepackage[all,cmtip]{xy}
\usepackage{setspace}
\usepackage{multirow}

\newcommand{\matr}[1]{\bm{\mathrm{#1}}}
\newcommand{\vect}[1]{\bm{#1}}
\newcommand{\transp}[0]{\! \mathsf{T}}

\title{Bayesian inference on dynamic linear models of day-to-day origin-destination flows in transportation networks}

\author{Anselmo Ramalho Pitombeira-Neto \footnote{Email:  anselmo.pitombeira@ufc.br.} \\
	    \small{Department of Industrial Engineering, Federal University of Ceará, Brazil} \\
	    Carlos Felipe Grangeiro Loureiro \\
	    \small{Department of Transportation Engineering, Federal University of Ceará, Brazil} \\
	    Luis Eduardo Carvalho \\ \small{Department of Mathematics and Statistics, Boston University}}
	
\date{}
	
\begin{document}
\maketitle

\begin{abstract}
	\noindent
	Estimation of origin-destination (OD) demand plays a key role in successful transportation studies. In this paper, we consider the estimation of time-varying day-to-day OD flows given data on traffic volumes in a transportation network for a sequence of days. We propose a dynamic linear model (DLM) in order to represent the stochastic evolution of OD flows over time. DLM's are Bayesian state-space models which can capture non-stationarity. We take into account the hierarchical relationships between the distribution of OD flows among routes and the assignment of traffic volumes on links. Route choice probabilities are obtained through a utility model based on past route costs. We propose a Markov chain Monte Carlo algorithm, which integrates Gibbs sampling and a forward filtering backward sampling technique, in order to approximate the joint posterior distribution of mean OD flows and parameters of the route choice model. Our approach can be applied to congested networks and in the case when data are available on only a subset of links. We illustrate the application of our approach through simulated experiments on a test network from the literature.
\end{abstract}


\pagebreak

\section{Introduction}

Given a geographic region, one of the main problems in planning and operating transportation systems is the estimation of origin-destination (OD) flows, i.e., the amount of trips made by people or freight between points in the region over a defined time interval. This is also referred to in the literature as the OD matrix estimation problem, since many initial models represented the vector of OD flows as a matrix.

OD flows are traditionally estimated through surveys, in which households or drivers are inquired about their daily journeys \cite{ortuzar}. However, direct surveys are expensive and due to this reason they are in general carried out every decade. \cite{cascetta2009}. This low frequency implies that planners may remain many years with no data on the evolution of OD flows over time.

In the last years, governments in urban regions have built   traffic control systems in order to manage traffic congestion in transportation networks. Those systems automatically gather large amounts of data on traffic volumes at low cost on a daily or even hourly basis. This allowed in theory the indirect estimation of OD flows by means of mathematical models. The idea is that we can extract information on OD flows from data on traffic volumes if we have  a suitable mathematical model which describes their relationships. These models are of an optimization or statistical nature.

The early models assumed that the transportation network from which one observes traffic volumes is not congested. We can trace back the first attempts to the work of \cite{robillard}, who applied linear regression to estimate OD flows from traffic volumes. \cite{nguyen77} proposed a model based on Beckman's optimization model for traffic assignment \cite{beckmann}, whose solution is an estimate of the OD flows.

\cite{vanzuylen80} proposed a non-linear programming model based on entropy maximization in which the constraints are the linear conservation flow equations, whose solution corresponds to a maximum entropy OD flow configuration. \cite{cascetta84} proposed a generalized least squares (GLS) model which can cope with errors in observed traffic volumes by minimizing the Mahalanobis distance between predicted volumes and observed volumes. \cite{cascetta88} and \cite{brenninger} proposed general frameworks which generalize previous models.

\cite{fisk88, fisk89} proposed models which take into account traffic congestion, by seeking solutions which correspond to an equilibrium state as defined for example by Wardrop's first principle \cite{wardrop1952, smith1979}. \cite{yang92} proposed a framework based on bilevel optimization, in which the first level corresponds to an objective function (e.g., maxent or GLS) subject to constraints which are obtained as a solution of a traffic assignment model in the second level. Due to the non-convex nature of the model, the authors proposed a heuristic procedure to solve it \cite{yang95}. 

\cite{postorino} formulated the OD matrix estimation problem as a compound fixed point problem, in which the solution of an inner fixed point problem corresponds to an equilibrium traffic assignment of an OD matrix obtained as a solution of an outer fixed point problem. In order to solve the problem, they propose a fixed point iteration based on the method of successive averages \cite{sheffi}.

A common caveat in the abovementioned models is that they do not take variability of OD flows into account. Assuming that OD flows follow independent Poisson probability distributions, \cite{vardi} considered the estimation of mean OD flows given a sample of observed traffic volumes vectors. Since the mean and variance of Poisson random variables is equal, variance of traffic volumes carry information on mean OD flows. He also assumed the existence of a single route for each OD pair, and proposed maximum likelihood and moment-based estimators for the mean OD flows.

Following Vardi's work, \cite{tebaldi} proposed Bayesian estimators for the case when a sample of size one is available for the traffic volumes vector. \cite{hazelton00, hazelton03, hazelton2001} relaxed the assumption of a single route between OD pairs and proposed multivariate normal approximations to the likelihood function assuming that the variances of OD flows are functions of their means.

More recent approaches consider the development of dynamic models for the estimation of OD flows. As traffic volumes are observed over time, we cannot ignore their temporal nature. Here we must make a distinction between \emph{within-day} and \emph{day-to-day} dynamics. When considering within-day dynamics, one is concerned with the estimation of OD flows over the course of a single day. Here we can cite the noteworthy works by \cite{cremer}, \cite{cascetta_dyn} and \cite{ben-akiva-dyn}. In contrast, in day-to-day dynamics we consider the estimation of time-varying OD flows for a reference time period, e.g. the morning peak, given a time series of observed traffic volumes over many consecutive days. It is assumed that this reference time period is long enough so that most trips in the study region start and finish within the observation period.

\cite{hazelton08} developed one of the first models for the estimation of day-to-day dynamic OD flows. The author assumes that mean OD flows are functions of parameters that do not change over time. Particular cases include constant demand, linear trend and weekday-weekend models. \cite{hazelton13} proposed a general Bayesian framework  for the estimation of parameters of day-to-day dynamic traffic models. They assumed that OD flows vary according to a Markovian transition kernel, and proposed a Markov chain Monte Carlo (MCMC) algorithm for the estimation of parameters. \cite{hazelton15} proposed statistical methods to compare alternative day-to-day dynamic models. 

A central aspect in the development of realistic day-to-day dynamic models is the modeling of users' behavior through route choice models. In practice, users choose routes based on the travel times they have experienced over past days. Thus, a suitable route choice model should take into account the influence of past route costs on the decision of users in a given time. \cite{watling2013, Cantarella2015} discuss models for the analysis of dynamic users' behavior.

In this paper, we represent stochastic day-to-day OD flows as a dynamic linear model (DLM), which allows us to capture the time dependencies of OD flows as well as non-stationarity. To the extent of our knowledge, this approach has not been applied before to the estimation of time-varying OD flows.  In the formulation of the DLM, we take into account the variability originating from OD flows, user's route choices and measurement of traffic volumes on links.  We model route choices through a utility model based on past route costs. Our model can be applied to congested networks and when there are available data only on partial links.

We also propose an MCMC algorithm, based on Gibbs sampling and forward filtering backward sampling, in order to approximate the joint posterior distributions of mean OD flows and parameters of the route choice model. We illustrate its application through numerical studies in a network from the literature. 

This paper is organized as follows: in Section \ref{sec:model} we describe the proposed dynamic linear model; in Section \ref{sec:route_choices} we describe a route choice model and an MCMC algorithm for the approximation of the joint posterior of mean OD flows and route choice parameters; in Section \ref{sec:numerical_studies} we present numerical studies and discuss the results; finally, in Section \ref{sec:conclusion} we draw some conclusions and propose further developments.

\section{A dynamic linear model for day-to-day OD flows}
\label{sec:model}

In Section \ref{sec:DLM} we define the proposed dynamic linear model and in Section \ref{sec:mean_OD_flows} we describe the  procedure for estimation of mean OD flows.

\subsection{Model definition}
\label{sec:DLM}

Dynamic linear models are defined by a state transition equation, which describes how a system evolves in time, and an observational equation, which describes how observed quantities relate to the system's states. They have a Markovian structure and assume multivariate normal probability distributions for the random variables. System's states are regarded as unobserved parameters whose estimation is done through Bayesian updating. We refer to \cite{west} for an excellent introduction to the corresponding theory.

Let $(\mathcal{N}, \mathcal{L})$ be a transportation network in which $\mathcal{N}$ is a set of nodes, $\mathcal{L}$ a set of directed links and $\mathcal{J} \subseteq \mathcal{N} \times \mathcal{N}$ a set of OD pairs. For a sequence of subsequent time periods $t = 1, 2, \dots, T$, we define $\bm{\theta}_t = (\theta_{1t}, \theta_{2t}, \dots, \theta_{nt})^{\transp}$ as the mean OD flow vector, in which $\theta_{jt}$ is the mean OD flow for OD pair $j \in \mathcal{J}$ at time $t$ and $n = |\mathcal{J}|$. We define $\vect{z}_t = (z_{1t}, z_{2t}, \dots, z_{mt})^{\transp}$ as the vector of observed traffic volumes in a subset of links in a network at time $t$, in which $z_{it}$ is the observed volume on link $i \in \mathcal{I} \subseteq \mathcal{L}$ and $m = |\mathcal{I}|$.

We consider the vector of mean OD flows $\bm{\theta}_t$ as the \emph{unobserved} system state vector, which depends  only on its previous state $\bm{\theta}_{t-1}$ at time $t-1$ plus a stochastic error. The vector of observed link volumes $\vect{z}_t$ at time $t$ depends only on the current unobserved vector of states, such that:

\begin{align}
\bm{\theta}_t & = \matr{G}_t \bm{\theta}_{t-1}+\bm{\omega}_t \label{eq:local_OD} \\
\vect{z}_t & = \matr{F}_t \bm{\theta}_t + \bm{\nu}_t \label{eq:observation_model}
\end{align}

\noindent We refer to equation \eqref{eq:local_OD} as the dynamic model, while equation \eqref{eq:observation_model} is the observational model. $\matr{G}_t$ is the system matrix and $\bm{\omega}_t \sim \mathrm{N}(\matr{0}, \matr{W}_t)$ is a stochastic error, where the letter ``N'' stands for the multivariate normal density with the appropriate dimension and $\matr{W}_t$ is a covariance matrix also referred to as an \emph{evolution matrix}. It governs the stochastic evolution of the system through time. We notice that $\bm{\theta}_t | \bm{\theta}_{t-1} \sim \mathrm{N}(\matr{G}_t\bm{\theta}_{t-1},\matr{W}_t)$. $\matr{F}_t$ is an \emph{assignment matrix} which relates mean OD flows to  observed traffic volumes, while $\bm{\nu}_t \sim \mathrm{N}(\vect{0}, \matr{V}_t)$ is a zero-mean error term which corresponds to the deviation of observed volumes relative to the expected value $\mathrm{E}[\vect{z}_t | \bm{\theta}_t] = \matr{F}_t \bm{\theta}_t$. We have that $\vect{z}_t | \bm{\theta}_t \sim \mathrm{N}(\matr{F}_t \bm{\theta}_t, \matr{V}_t)$.

In order to fully determine the model given by equations \eqref{eq:local_OD} and \eqref{eq:observation_model}, we must specify the corresponding matrices. Regarding the system matrix, we assume that in the short term mean OD flows are \emph{locally constant}. This implies that at time $t$ mean OD flows should be approximately equal to mean OD flows at time $t-1$, so that $\bm{\theta}_t = \bm{\theta}_{t-1}+\bm{\omega}_t$ and $\matr{G}_t = \matr{I}$, the identity matrix with the appropriate dimension. The evolution matrix $\matr{W}_t$ is in general supplied by the modeler.

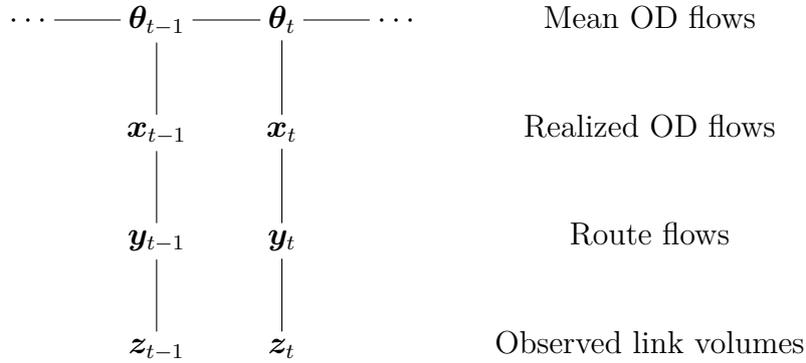
\begin{figure}[htbp]
	\[
	\xymatrix{
		\cdots \ar@{-}[r] & \bm{\theta}_{t-1} \ar@{-}[r] \ar@{-}[d] &
		\bm{\theta}_t \ar@{-}[r] \ar@{-}[d] & \cdots &
		\text{Mean OD flows} \\
		& \vect{x}_{t-1} \ar@{-}[d] & \vect{x}_t \ar@{-}[d] &  &
		\text{Realized OD flows} \\
		& \vect{y}_{t-1} \ar@{-}[d] & \vect{y}_t \ar@{-}[d] &  &
		\text{Route flows} \\
		& \vect{z}_{t-1}            & \vect{z}_t            &  &
		\text{Observed link volumes}
	}
	\]
	\caption{Hierarchical relationship between variables in the DLM. Mean OD flows, realized OD flows and route flows are unobserved variables.}
	\label{fig:diagram}
\end{figure}

Regarding matrices $\matr{F}_t$ and $\matr{V}_t$ of the observational model, they must represent correctly the relationship between mean OD flows and observed traffic volumes. We define $\vect{x}_t$ as \emph{realized} OD flows around mean OD flows at time $t$, which distributes among flows on routes $\vect{y}_t$, which aggregate into observed traffic volumes $\vect{z}_t$. Figure \ref{fig:diagram} summarizes the hierarchical relationship between variables in our proposed DLM.

Let $\vect{x}_t = (x_{1t}, x_{2t}, \dots, x_{nt})^{\transp}$ be the vector of realized OD flows at time $t$. We assume $\vect{x}_t | \bm{\theta}_t \sim \mathrm{N}(\bm{\theta}_t, \bm{\Sigma}_t^x)$ where $\bm{\Sigma}_t^x$ is a covariance matrix, which can account for correlations between OD flows or simply be a diagonal matrix in case OD flows are independent. Given a realized vector $\vect{x}_t$, for each OD pair $j$ there is a vector of route flows $\vect{y}_{jt} = (y_{1jt}, y_{2jt}, \dots, y_{n(j)jt })$, in which $n(j) = |\mathcal{K}_j|$ is the size of the route set $\mathcal{K}_j$ of OD pair $j$, and let $p_{kjt}$ be the probability of route $k \in \mathcal{K}_j$ being selected at time $t$. We assume $\vect{y}_{jt}$ follows a multivariate normal distribution with a multinomial-like covariance structure, i.e.:

\begin{align*}
\mathrm{E}[y_{kjt}] & = x_{jt} p_{kjt} \\
\mathrm{Cov}(y_{kjt}, y_{ljt}) & =		
\begin{cases}
 x_{jt} p_{kjt} (1-p_{kjt}) & \text{if} \quad  l = k \\
- x_{jt}  p_{kjt}p_{ljt} & \text{if} \quad l \neq k
\end{cases}
\end{align*}

\noindent
Where $\mathrm{E[.]}$ and $\mathrm{Cov}(.)$ denote the expected value and the covariance, respectively. Thus $\vect{y}_{jt} |x_{jt}, \vect{p}_{jt} \sim \mathrm{N}(x_{jt}\vect{p}_{jt}, \bm{\Sigma}_{jt}^y)$, in which $\vect{p}_{jt} = (p_{1jt}, p_{2jt}, \dots, p_{n(j)jt})^{\transp}$ is the vector of route choice probabilities of OD pair $j$ and the covariance matrix is given by:

\begin{equation}
\bm{\Sigma}_{jt}^{y} = x_{jt}(\mathrm{diag}(\vect{p}_{jt}) - \vect{p}_{jt}\vect{p}_{jt}^{\transp}) \label{eq:Sigma_routes}
\end{equation}

\noindent
Where $\mathrm{diag}(\vect{p}_{jt})$ denotes a matrix whose diagonal elements correspond to the vector $\vect{p}_{jt}$ and all other elements are zero. Notice also that as $\sum_{k \in \mathcal{K}_j} p_{kjt} = 1$, the covariance matrix $\bm{\Sigma}_{jt}^{y}$ will be singular and the corresponding multivariate normal distribution will be degenerate. In order to avoid this, we assume there is a positive probability $\pi$ of none of the routes in a route set being chosen, so that $\sum_{k \in \mathcal{K}_j} p_{kjt} < 1$ assuring that $\bm{\Sigma}_{jt}^{y}$ is non-singular. This assumption will be realistic if the route sets include many routes, so that flows on the excluded routes are negligible.

Let $\vect{y}_t = (\vect{y}_{1t}, \vect{y}_{2t}, \dots, \vect{y}_{nt})^{\transp}$ be the joint vector of route flows across all OD pairs. We also define $\matr{P}_t = \mathrm{blockdiag}_{j \in \mathcal{J}} \lbrace \vect{p}_{jt} \rbrace$ as a block-diagonal matrix composed of route choice probability vectors and $\bm{\Sigma}_t^y = \mathrm{blockdiag}_{j \in \mathcal{J}} \lbrace \bm{\Sigma}_{jt}^{y} \rbrace$ as a block-diagonal matrix composed of covariance matrices of route flows, i.e.:

\begin{align}
\matr{P}_t & = \begin{bmatrix}
\vect{p}_{1t} & & & \\
& \vect{p}_{2t} & & \\
& & \ddots & \\
& & & \vect{p}_{nt}
\end{bmatrix} \label{eq:route_choice_matrix} \\ \nonumber \\ 
\bm{\Sigma}_t^y  &= \begin{bmatrix}
\bm{\Sigma}_{1t}^y & & & \\
& \bm{\Sigma}_{2t}^y & & \\
& & \ddots & \\
& & & \bm{\Sigma}_{nt}^y
\end{bmatrix}
\end{align}

\noindent
Since we defined multivariate normal distributions for all subvectors $\vect{y}_{jt}$, and assuming that route flow subvectors $\vect{y}_{jt}$ are conditionally independent given realized OD flow vector $\vect{x}_t$ we have that $\vect{y}_t |\vect{x}_t, \matr{P}_t  \sim \mathrm{N}(\matr{P}_t \vect{x}_t, \bm{\Sigma}_t^y)$. (For convenience, we omit the explicit dependence on $\matr{P}_t$ in the following developments).

Now we are able to obtain the conditional $\mathrm{p}(\vect{y}_t | \bm{\theta}_t)$ by marginalizing $\vect{x}_t$. Since both the densities $p(\vect{x}_t | \bm{\theta}_t)$ and $p(\vect{y}_t | \vect{x}_t)$ are multivariate normal, and ignoring the dependence of $\bm{\Sigma}_t^y$ on $\vect{x}_t$ (we will take the dependence of $\bm{\Sigma}_t^y$ on $\vect{x}_t$ into account in the inference of $\bm{\theta}_t$ in next section), from the properties of the multivariate normal we have \cite{sarkka}:

\begin{equation*}
p(\vect{y}_t | \bm{\theta}_t) = \mathrm{N}(\matr{P}_t \bm{\theta}_t, \matr{P}_t \bm{\Sigma}_t^x \matr{P}^{\transp}+ \bm{\Sigma}_t^y) \label{eq:marginal_routes_2}
\end{equation*}

In order to complete the formulation of our model, we must obtain the conditional density $p(\vect{z}_t | \bm{\theta}_t)$ of observed traffic volumes given mean OD flows. First we assume that $\vect{z}_t | \vect{y}_t \sim \mathrm{N}(\bm{\Delta}\vect{y}_t, \bm{\Sigma}_t^z)$,  where $\bm{\Sigma}_t^z$ is the covariance matrix of measurement errors when observing volumes on links. $\bm{\Delta}$ is the link-path incidence matrix for observed links, whose entries $a_{ik} = 1$ if link $i$ make part of route $k$ and $a_{ik} = 0$ otherwise. It is a deterministic parameter which is a function of the topology of the network and of the route choice sets. 

Since both $p(\vect{z}_t | \vect{y}_t)$ and $p(\vect{y}_t |\bm{\theta}_t)$ are multivariate normal, by marginalizing $\vect{y}_t$ we have:

\begin{equation}
p(\vect{z}_t | \bm{\theta}_t) = \mathrm{N}(\bm{\Delta}\matr{P}_t \bm{\theta}_t, \bm{\Delta}(\matr{P} \bm{\Sigma}_t^x\matr{P}_t^{\transp}+ \bm{\Sigma}_t^y)\bm{\Delta}^{\transp}+\bm{\Sigma}_t^z) \label{eq:dist_z}
\end{equation}

\noindent
From \eqref{eq:dist_z}, we can finally specify $\matr{F}_t$ and $\matr{V}_t$ by defining:

\begin{equation}
    \matr{F}_t = \bm{\Delta}\matr{P}_t \label{eq:assignment}
\end{equation}

\noindent
as an assignment matrix and the covariance matrix by:  

\begin{equation}
\matr{V}_t = \matr{F}_t \bm{\Sigma}_t^x\matr{F}_t^{\transp}+\bm{\Delta} \bm{\Sigma}_t^y\bm{\Delta}^{\transp}+\bm{\Sigma}_t^z \label{eq:V}
\end{equation}

\noindent
It is noteworthy in equation \eqref{eq:V} that the variability in OD flows, in route choices and in volume measurement are all represented in the covariance matrix.

In next section we describe the estimation equations for mean OD flows.

\subsection{Estimation of mean OD flows}
\label{sec:mean_OD_flows}

Estimation of mean OD flows $\bm{\theta}_t$ from observed link volumes can be made by Bayesian updating. We assume that the covariance matrices $\bm{\Sigma}_t^x$, $\bm{\Sigma}_t^z$ and $\matr{W}_t$, the link-path incidence matrix $\bm{\Delta}$, the route choice matrices $\matr{P}_t$ and the probability $\pi$ of none of the routes in a route set being chosen are known parameters and make part of prior knowledge.

At time $t$, let $p(\bm{\theta}_{t-1} | \vect{z}_{0:t-1}) = \mathrm{N}(\vect{m}_{t-1}, \matr{C}_{t-1})$ be the posterior distribution in previous time step $t-1$, where $\vect{z}_{0:t-1} = \lbrace \vect{z}_0, \vect{z}_1, \dots, \vect{z}_{t-1} \rbrace$ is the set of observed volume vectors in past time periods. We have a prior distribution $p(\bm{\theta}_t | \vect{z}_{0:t-1}) = \mathrm{N}(\bar{\vect{m}}_{t}, \bar{\matr{C}}_{t})$ before the observation of $\vect{z}_t$. Since in our local constant model given by equation \eqref{eq:local_OD} the system matrix $\matr{G}_t = \matr{I}$, then:

\begin{align}
    \bar{\vect{m}}_t & = \vect{m}_{t-1} \label{eq:prior_mean} \\ 
    \bar{\matr{C}}_{t} & = \matr{C}_{t-1}+\matr{W}_t \label{eq:prior_cov}
\end{align}

\noindent
Notice that uncertainty on mean OD flows increases by addition of the evolution matrix $\matr{W}_t$ to the covariance matrix $\matr{C}_{t-1}$. Next, we define $p(\vect{z}_{t}|\vect{z}_{0:t-1}) = \mathrm{N}(\vect{f}_{t}, \matr{Q}_{t})$ as the one-step forecast distribution of the vector of observed link volumes, where:

\begin{align}
    \vect{f}_{t} & = \matr{F}_{t} \bar{\vect{m}}_{t} \label{eq:one_step_mean}\\
    \matr{Q}_{t} & = \matr{F}_{t} \bar{\matr{C}}_{t} \matr{F}_{t}^{\transp} + \matr{V}_{t} \label{eq:one_step_cov}
\end{align}

\noindent
It is worth noting that in computing $\matr{V}_{t}$ by equation \eqref{eq:V}, the covariance matrix of route flows $\bm{\Sigma}_t^y$ is computed based on the predicted mean OD flows $\bar{\vect{m}}_t = (\bar{m}_{1t}, \bar{m}_{2t}, \dots, \bar{m}_{nt})^{\transp}$, i.e., $\bm{\hat{\Sigma}}_t^y = \mathrm{blockdiag}_{j \in \mathcal{J}} \lbrace \bm{\hat{\Sigma}}_{jt}^{y} \rbrace$ in which from equation \eqref{eq:Sigma_routes}:

\begin{equation}
\bm{\hat{\Sigma}}_{jt}^{y} = \bar{m}_{jt}(\mathrm{diag}(\vect{p}_{jt}) - \vect{p}_{jt}\vect{p}_{jt}^{\transp})   \label{eq:cov_routes_estimated}
\end{equation}

Finally, the parameters of the posterior distribution $\mathrm{p}(\bm{\theta}_t | \vect{z}_{0:t}) = \mathrm{N}(\vect{m}_t, \matr{C}_t)$ are given by the Bayesian updating of the parameters, which in this case coincide with Kalman filter updating equations \cite{west}:

\begin{align}
	\vect{m}_t & =  \bar{\vect{m}}_t+ \matr{A}_t (\vect{z}_t - \vect{f}_t) \label{eq:update_mean} \\
	\matr{C}_t & =  \bar{\matr{C}}_t - \matr{A}_t \matr{Q}_t \matr{A}_t^{\transp} \label{eq:update_covariance} \\
	\matr{A}_t & = \bar{\matr{C}}_t\matr{F}_t^{\transp}\matr{Q}_t^{-1} \label{eq:adjustment_matrix}
\end{align}

\noindent
Where $\matr{A}_t$ is an adjustment matrix, which controls how the parameters from the posterior distribution are modified according to the new observation $\vect{z}_t$. In particular, the adjustment matrix is a function of the prior covariance matrix $\bar{\matr{C}_t}$ and of the inverse $\matr{Q}^{-1}_t$ of the covariance matrix of the one-step forecast distribution of link volumes, so that the adjustment matrix gives more or less weight to observed link volumes according to their uncertainty relative to the uncertainty in OD flows.

At a time $t$, $\bm{\hat{\theta}}_t = \vect{m}_t$ is an estimator of the mean OD flows and $\matr{C}_t$ is a measure of uncertainty of the estimate. At $t = 0$, we define $p(\bm{\theta}_0 | \vect{z}_0) = \mathrm{N}(\vect{m}_0, \matr{C}_0)$ where $\vect{z}_0$ are not actually observed link volumes, but symbolically represents the modeler's prior knowledge on OD flows. The estimation procedure can be summarized as follows:

\begin{enumerate}
	\item Starting from $t = 0$, set $p(\bm{\theta}_0 | \vect{z}_0) = \mathrm{N}(\vect{m}_0, \matr{C}_0)$;
	\item For $t=1$ to $T$, do:
	\begin{enumerate}
		\item Determine prior distribution parameters $\bar{\vect{m}}_t$ and $\bar{\matr{C}}_{t}$ by means of equations \eqref{eq:prior_mean} and \eqref{eq:prior_cov}, respectively;
		\item Compute the assignment matrix $\matr{F}_t$ by means of equation \eqref{eq:assignment}
		\item Compute the covariance matrix $\matr{V}_t$ by means of equations \eqref{eq:V} and \eqref{eq:cov_routes_estimated};
		\item Determine the parameters of the one-step forecast distribution $\vect{f}_{t}$ and $\matr{Q}_{t}$ by using equations \eqref{eq:one_step_mean} and \eqref{eq:one_step_cov};
		\item Determine posterior distribution parameters $\vect{m}_t$ and $\matr{C}_t$ by means of equations \eqref{eq:update_mean}, \eqref{eq:update_covariance} and \eqref{eq:adjustment_matrix}.
	\end{enumerate}
\end{enumerate}

In next section, we model route choice probabilities based on a utility model in order to estimate jointly mean OD flows and the route choice matrix.

\section{Bayesian inference on route choice parameters}
\label{sec:route_choices}

In Section \ref{sec:route_choice}, we treat route choice probabilities as unobserved quantities and represent them through a utility model. In Section \ref{sec:Gibbs_sampler} we propose an MCMC algorithm to approximate the joint posterior distribution of route choice parameters and mean OD flows.

\subsection{Utility model and route choice probabilities}
\label{sec:route_choice}

In congested networks, route choice probabilities greatly depend on past route costs. Users take into account their previous experiences in order to evaluate the utility associated with each route. Utilities will vary over time depending on user's memory, their sensitivity to route costs, fluctuations in mean OD flows, among other possibly influencing factors. As route choice probabilities are functions of utilities, these will also dynamically change.

We assume that the utility $u_{kjt}$ of a route $k$ in an OD pair $j$ at a time $t$ as perceived by users is a linear function of its past route costs:

\begin{equation}
    u_{kjt} = - \sum_{s = 1}^{r}\phi_s c_{kj(t-s)} \label{eq:utility}
\end{equation}

\noindent
in which $\phi_{s}$ are parameters, $c_{kj(t-s)}$ are observed past route costs and $r$ is the length of users' memory. The minus sign in equation \eqref{eq:utility} is used since costs are disutilities, in order that routes with lower costs are preferable. We may interpret the parameters $\bm{\phi} = (\phi_1, \phi_2, \dots, \phi_r)^{\transp}$ as users' \emph{sensitivities} to past route costs. It is reasonable to assume that they have non-negative values, otherwise higher costs will contribute to higher utilities. We also expect users to be more sensitive to recent route costs than to older costs, which implies that $\phi_1 \geq \phi_2 \geq, \dots, \geq \phi_r$.  

We adopt a multinomial logit model for route choice probabilities \cite{ben-akiva} in order that the probability $p_{kjt}$ of a route $k \in \mathcal{K}_j$ in OD pair $j$ at time $t$ being chosen by a user is given by:

\begin{equation}
    p_{kjt} = (1-\pi) \frac{e^{u_{kjt}}}{\sum_{l \in \mathcal{K}_j} e^{u_{ljt}}}  \label{eq:logit}
\end{equation}

\noindent
Where $\pi$ is the probability of a user not following a route in a route choice set and $e$ denotes Euler's number.

Let $\vect{c}_{jt} = (c_{1jt}, c_{2jt}, \dots,  c_{n(j)jt})^{\transp}$ be the vector of route costs for routes in OD pair $j$ at time $t$ and $\vect{c}_t = (\vect{c}_{1t}, \vect{c}_{2t}, \dots, \vect{c}_{nt})^{\transp}$. We notice that, given route costs $\vect{c}_t$ for $t = -(r-1), ..., 0, 1, \dots, T$, and parameters $\bm{\phi}$ and $\pi$, we uniquely determine the vectors $\vect{p}_{jt} = (p_{1jt}, p_{2jt}, \dots, p_{n(j)jt})^{\transp}$ and the route choice matrix $\matr{P}_t$ for $t=1$ to $T$ by means of equations \eqref{eq:route_choice_matrix},  \eqref{eq:utility} and \eqref{eq:logit}.

\subsection{An MCMC algorithm}
\label{sec:Gibbs_sampler}

In this section, we develop an MCMC algorithm in order to sample from the joint posterior distribution of mean OD flows and the parameters of the route choice utility model given in \eqref{eq:utility}.

From Bayes theorem, we have:

\begin{equation}
p(\bm{\theta}_{1:T}, \bm{\phi} | \vect{z}_{1:T}, D_0) \propto  p(\vect{z}_{1:T} | \bm{\theta}_{1:T}, \bm{\phi}, D_0) p(\bm{\theta}_{1:T}, \bm{\phi} | D_0) \label{eq:joint_posterior}
\end{equation}

\noindent
where $\bm{\theta}_{1:T}$ is a shorthand referring to the vectors $\bm{\theta}_1, \bm{\theta}_2, \dots, \bm{\theta}_T$ and $\vect{z}_{1:T}$ a shorthand for $\vect{z}_1, \vect{z}_1, \dots, \vect{z}_T$. $p(\bm{\theta}_{1:T}, \bm{\phi} | \vect{z}_{1:T}, D_0)$ is the joint posterior, $p(\vect{z}_{1:T} | \bm{\theta}_{1:T}, \bm{\phi}, D_0)$ is the likelihood function of the observed link volumes and $p(\bm{\theta}_{1:T}, \bm{\phi} | D_0)$ is the prior distribution. We include in prior knowledge set $D_0$ all other parameters, such as $\pi, \bm{\Delta}, \matr{W}_t, \bm{\Sigma}_t^x, \bm{\Sigma}_t^z$ for $t = 1, \dots, T$, and route costs $\vect{c}_t$ for $t = -(r-1), ..., 0, 1, \dots, T$.

We notice that Gibbs sampling \cite{geman} allows us to sample from the joint distribution $p(\bm{\theta}_{1:T}, \bm{\phi} | \vect{z}_{1:T}, D_0)$ by alternately sampling from conditional distributions $p(\bm{\theta}_{1:T} | \bm{\phi}, \vect{z}_{1:T}, D_0)$ and $p(\bm{\phi} | \bm{\theta}_{1:T}, \vect{z}_{1:T}, D_0)$. We must then devise how to sample from each distribution. 

Sampling from $p(\bm{\theta}_{1:T} | \bm{\phi}, \vect{z}_{1:T}, D_0)$ can be implemented through the \emph{forward filtering backward sampling} (FFBS) method \cite{carter_kohn, schnatter, west}:

\begin{enumerate}
	\item (forward filtering) For $t = 1, \dots, T$, apply Kalman filtering recurrences given by equations \eqref{eq:update_mean} and \eqref{eq:update_covariance}. Save $\vect{m}_t$, $\bar{\vect{m}}_t$, $\matr{C}_t$ and $\bar{\matr{C}}_t$ for $t = 1, \dots, T$;
	\item (backward sampling) Starting from $\bm{\theta}_T \sim \mathrm{N}(\vect{m}_t, \matr{C}_t)$, for $t = T-1, \dots, 1$, sample each $\bm{\theta}_t$ backwards from the conditionals $p(\bm{\theta}_t | \bm{\theta}_{t+1}, \bm{\phi}, \vect{z}_{1:T}) = \mathrm{N}(\vect{h}_t, \matr{H}_t)$, in which:
	
	\begin{align}
	\vect{h}_t & = \vect{m}_t + \matr{B}_t (\bm{\theta}_{t+1} - \bar{\vect{m}}_{t+1}) \label{eq:FFBS_mean}\\ 
	\matr{H}_t & = \matr{C}_t-\matr{B}_t \bar{\matr{C}}_t \matr{B}_t^{\transp} \label{eq:FFBS_cov}
	\end{align}

	and $\matr{B}_t = \matr{C}_t \bar{\matr{C}}_{t+1}^{-1}$.
\end{enumerate}

We notice that step 1 (forward filtering) of the FFBS algorithm is possible since given $\bm{\phi}$, observed link volumes $\vect{z}_t$ and prior knowledge $D_0$, we have the route choice matrix $\matr{P}_t$ and all other parameters determined for $t = 1, \dots, T$, so that we can apply recurrence equations \eqref{eq:update_mean} and \eqref{eq:update_covariance}. Then, step 2 generates a sample of mean OD flows $\bm{\theta}_{1:T}$.

Regarding the conditional $p(\bm{\phi} | \bm{\theta}_{1:T}, \vect{z}_{1:T}, D_0)$, sampling can be performed through a Metropolis-Hastings (MH) step \cite{hastings}. First, we notice that:

\begin{align}
    p(\bm{\phi} | \bm{\theta}_{1:T}, \vect{z}_{1:T}, D_0) & \propto
    p(\vect{z}_{1:T} | \bm{\phi}, \bm{\theta}_{1:T}, D_0) p(\bm{\theta}_{1:T} | \bm{\phi}, D_0) p(\bm{\phi} | D_0) \nonumber \\
    & \propto p(\vect{z}_{1:T} | \bm{\phi}, \bm{\theta}_{1:T}, D_0) p(\bm{\phi} | D_0) \label{eq:phi_posterior}
\end{align}

\noindent
as $\bm{\theta}_{1:T}$ does not depend on $\bm{\phi}$, and $p(\bm{\phi} | D_0)$ is the prior distribution of $\bm{\phi}$. From \eqref{eq:dist_z}, the likelihood term is given by:

\begin{equation}
    p(\vect{z}_{1:T} | \bm{\phi}, \bm{\theta}_{1:T}, D_0) \propto \prod_{t = 1}^{T} {||\matr{V}_t||}^{-1/2} \exp \left\lbrace {-\frac{1}{2} (\vect{z}_t - \matr{F}_t \bm{\theta}_t)^{\transp} \matr{V}_t^{-1} (\vect{z}_t - \matr{F}_t \bm{\theta}_t)} \right\rbrace
\end{equation}

\noindent
in which matrices $\matr{F}_t$ and $\matr{V}_t$ are dependent on $\bm{\phi}$ since they are functions of the route choice matrices $\matr{P}_t$, which are dependent on $\bm{\phi}$ through equations \eqref{eq:utility} and \eqref{eq:logit}.

The MH step is then performed as follows: given a candidate vector $\bm{\phi}^{\prime} \sim q(\bm{\phi}^{\prime} | \bm{\phi}^{(i)})$, where $\bm{\phi}^{(i)}$ is the current vector at iteration $i$ and $q(.)$ is a proposal distribution, sample $u \sim \mathrm{Uniform}(0,1)$ and accept $\bm{\phi}^{\prime}$ as next sample $\bm{\phi}^{(i+1)}$  if:

\begin{equation}
u < \min \left(1, \frac{p(\vect{z}_{1:T} | \bm{\phi}^{\prime}, \bm{\theta}_{1:T}, D_0) p(\bm{\phi}^{\prime} | D_0) q( \bm{\phi}^{(i)} |\bm{\phi}^{\prime} )}{p(\vect{z}_{1:T} | \bm{\phi}^{(i)}, \bm{\theta}_{1:T}, D_0) p(\bm{\phi}^{(i)} | D_0) q(\bm{\phi}^{\prime} | \bm{\phi}^{(i)})} \right) \label{eq:acceptance_test}
\end{equation}

\noindent
otherwise, make $\bm{\phi}^{(i+1)} = \bm{\phi}^{(i)}$.

Finally, the proposed MCMC algorithm is summarized below:

\begin{enumerate}
	\item Initialize vectors $\bm{\phi}^{(0)}$ and $\bm{\theta}_{1:T}^{(0)}$.
	\item From iteration $i = 0$ onwards, repeat until convergence:
	\begin{enumerate}
		\item (FFBS step) Sample $\bm{\theta}_{1:T}^{(i+1)}$ by applying updating equations \eqref{eq:update_mean}, \eqref{eq:update_covariance} and sampling backwards through equations \eqref{eq:FFBS_mean} and \eqref{eq:FFBS_cov};
		\item (MH step) Sample a candidate $\bm{\phi}^{\prime}$ according to a proposal distribution $q(\bm{\phi}^{\prime} | \bm{\phi}^{(i)})$ and make $\bm{\phi}^{(i+1)} = \bm{\phi}^{\prime}$ according to acceptance test given in equation \eqref{eq:acceptance_test}, otherwise make $\bm{\phi}^{(i+1)} = \bm{\phi}^{(i)}$.
	\end{enumerate}
\end{enumerate}

In next section we illustrate the application of the proposed MCMC algorithm through some numerical studies.

\section{Numerical studies}
\label{sec:numerical_studies}

In the following subsections, we describe the generation of simulated data and present the results from the application of our approach to a test network.

\subsection{Generation of simulated data}

\begin{figure}[h]
	\centering
	\includegraphics[trim={0cm, 3.0cm, 0cm, 2.0cm}, clip, scale=0.7]{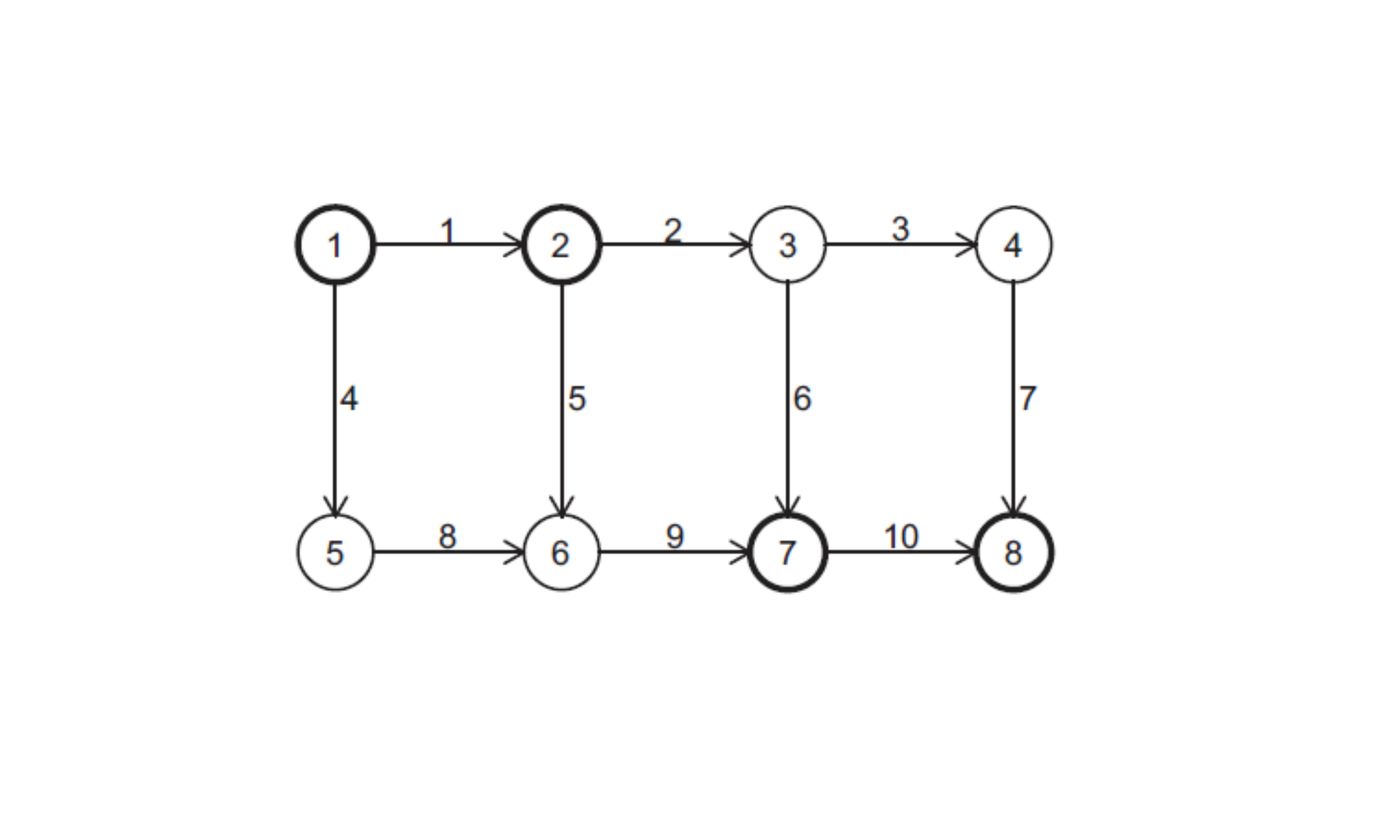}
	\caption{Network used in the numerical studies \cite[adapted from][]{hazelton15}}
	\label{fig:net_hazelton_2015}
\end{figure}

In order to illustrate the application of our proposed DLM and MCMC algorithm, we generated data for a transportation network from \cite{hazelton15}, given in Figure \ref{fig:net_hazelton_2015}, which has 8 nodes and 10 links. Nodes 1 and 2 are the origins and nodes 7 and 8 are the destinations, so that we have 4 OD pairs: (1,7), (1,8), (2,7) and (2,8).

In order to adapt the network to our experiment, we did not use the original data on demand and link parameters. We assumed that travel times on links are given by BPR functions according to equation \eqref{eq:bpr}.

\begin{equation}
\tau(z)=\tau_{0} \left[1+\alpha \left(\frac{z}{z^{\mathrm{max}}}\right)^\beta \right] \label{eq:bpr}
\end{equation}

Where $\tau(z)$ is the travel time on a link with traffic volume $z$, $\tau_{0}$ denotes the travel time in ``free flow'' and $z^{\mathrm {max}}$ is the capacity of the link. $\alpha$ and $\beta$ are parameters of the function, for which we adopt the typical values from the literature 0.15 and 4, respectively. Note that the capacity of the link is not treated as a hard constraint, i.e., simulated volumes $z$ are allowed to be greater than $z^{\mathrm{max}}$. In the test network, we set $\tau_{0} = 1$ and $z^{\mathrm {max}} = 130$ on all links. 

The procedure used to generate the simulated data was the following:

\begin{enumerate}
	\item At $t = 0$, set values for $\bm{\Delta}$, $\bm{\theta}_0$, $\matr{W}_t$, $\bm{\Sigma}_t^x$, $\bm{\Sigma}_t^z$, $\bm{\phi}$, $\pi$ for $t=1$ to $T$ and past route costs $\vect{c}_{1-r} = \vect{c}_{1-r+1} = \dots = \vect{c}_0 = \vect{c}_\mathrm{free flow}$, where $r$ is the size of users' memory and $\vect{c}_\mathrm{free flow}$ denotes route costs computed assuming free flow in the network.
	\item For $t = 1$ to $T$ do:
	\begin{enumerate}
		\item Sample mean OD flows $\bm{\theta}_t \sim \mathrm{N}(\bm{\theta}_{t-1}, \matr{W}_t)$;
		\item Sample OD flows $\vect{x}_t | \bm{\theta}_t \sim \mathrm{N}(\bm{\theta}_t, \bm{\Sigma}_t^x);$
		\item Compute $u_{kt}$ and $p_{kt}$ for all routes $k \in \mathcal{K}_j$ for all OD pairs $j \in \mathcal{J}$ from equations \eqref{eq:utility} and \eqref{eq:logit}, respectively, and compute the route choice matrix $\matr{P}_t = \mathrm{blockdiag}_{j \in \mathcal{J}} \lbrace \vect{p}_{jt} \rbrace$;
		\item Sample route flows $\vect{y}_t |\vect{x}_t \sim \mathrm{N}(\matr{P}_t \vect{x}_t, \bm{\Sigma}_t^y)$, where $\bm{\Sigma}_t^y$ is calculated from equation \eqref{eq:Sigma_routes};
		\item Calculate route costs $\vect{c}_t = \bm{\Delta}^{\transp}g(\bm{\Delta}\vect{y}_t)$, where $g(.)$ is a vector function which returns a vector of costs on links based on BPR functions given by equation \eqref{eq:bpr};
		\item Sample observed traffic volumes $\vect{z}_t| \vect{y}_t \sim \mathrm{N}(\bm{\Delta}\vect{y}_t, \bm{\Sigma}_t^z)$.
	\end{enumerate}
\end{enumerate}

OD flows vary within the interval $[10.0, 100.0]$ and follow the locally constant model given in equation \eqref{eq:local_OD}, starting at time $ t = 0$ from the OD flow vector $\bm{\theta}_0 = (50.0, 50.0, 50.0, 50.0)^{\transp}$. The OD pairs are ordered lexicographically in vector $\bm{\theta}_t$, so that mean OD flows in OD pair (1,7) corresponds to $\theta_{1t}$, (1,8) corresponds to $\theta_{2t}$ and so on. The evolution covariance matrix is constant and given by $\matr{W}_t = 10 \matr{I}$, where $\matr{I}$ is the appropriate identity matrix. We assume that variability around mean OD flows is negligible, so as to $\bm{\Sigma}_t^x = \matr{I}$ and $\bm{\theta}_t \approx \vect{x}_t$ for all $t$, and we also assume that $\bm{\Sigma}_t^z = \matr{I}$. These assumptions imply that variability in observed link volumes will be mostly due to variability in route choices. We simulate OD flows for $T = 100$ time periods.

In addition, we set a users' memory length of $r=2$ for the utility model, so that utility of a route $k$ of OD pari $j$ at time $t$ is given by $u_{kjt} = - \phi_1 c_{kj(t-1)} - \phi_2 c_{kj(t-2)}$ (equation \eqref{eq:utility}) and $\bm{\phi} = (\phi_1, \phi_2)^{\transp} = (0.5, 0.3)^{\transp}$ is the vector of route choice parameters. We enumerated exhaustively all 12 routes between origins and destinations, resulting in a link-path incidence matrix $\bm{\Delta}$ with 10 rows and 12 columns. Since we enumerated all routes, we assumed a small probability $\pi = 0.01$ of a trip not following one of the routes. Table \ref{tab:congestion} shows the mean congestion level (CL) for a link $i$, given by the ratio between mean simulated traffic volumes and the link capacities:

\begin{equation}
\mathrm{CL_i} = \sum_{t = 1}^{T} z_{it} /Tz^{\mathrm{max}}
\end{equation}

\noindent
As can be seen in Table \ref{tab:congestion}, links 2, 5, 6 and 9, which are in the central area of the network, have high congestion levels.

\begin{table}[h]
	\centering
	\caption{Mean congestion level (CL) on links from simulated OD flows.}
	\label{tab:congestion}
	\begin{tabular}{llllll}
		\hline
		Link & \multicolumn{1}{c}{1} & \multicolumn{1}{c}{2} & \multicolumn{1}{c}{3} & \multicolumn{1}{c}{4} & \multicolumn{1}{c}{5}  \\ \hline
		CL   & 0.5500                & 0.8940                & 0.2482                & 0.2535                & 0.6378                 \\ \hline
		Link & \multicolumn{1}{c}{6} & \multicolumn{1}{c}{7} & \multicolumn{1}{c}{8} & \multicolumn{1}{c}{9} & \multicolumn{1}{c}{10} \\ \hline
		CL   & 0.6473                & 0.2481                & 0.2542                & 0.8914                & 0.5685                 \\ \hline
	\end{tabular}
\end{table}

The simulations, the proposed DLM and MCMC algorithm were implemented in the Python programming language version 2.7 by using Numerical Python module version 1.10.

\subsection{Application of the MCMC algorithm}
\label{sec:application}

When applying the MCMC algorithm, we assume that we observe link volumes $\vect{z}_{1:T}$ on all links and route costs $\vect{c}_{1:T}$.  In practice, these data on traffic volumes are routinely collected through modern ITS systems and route costs can be estimated from data on travel times on links or can be collected from a sample of cars following selected routes. Except for the mean OD flows $\bm{\theta}_t$ and the parameters $\bm{\phi}$, we assume all other parameters are known. We use an uninformative multivariate normal prior for the initial mean OD flows vector $\bm{\theta}_0$, with $\vect{m}_0 = 100 \, \vect{1}$ where $\vect{1}$ is the unit vector with the appropriate dimension and $\matr{C}_0 = 1000 \, \matr{I}$, and an uninformative improper prior for $\bm{\phi}$.

The MCMC algorithm was ran through 10000 iterations with a multivariate normal proposal distribution for candidate $\bm{\phi}$ vectors with covariance matrix equals to $0.04 \, \matr{I}$, which resulted in an acceptance percentage of about 22\%, and we discarded the initial 2000 samples as burn-in. We used as starting values for $\bm{\phi}^{(0)} = (1.0, 1.0)^{\transp}$ and the starting values for $\bm{\theta}_{1:T}^{(0)}$ are sampled from a multivariate normal with mean $100 \, \vect{1}$ and covariance matrix $100 \, \matr{I}$. It took about 10 min to run the MCMC algorithm in a Core i7 machine with 3.1 GHz and 8GB RAM.

\begin{figure}[h!]
	\centering
	\includegraphics[trim={3.0cm, 1.0cm, 3.0cm, 1.0cm}, clip, scale=0.30]{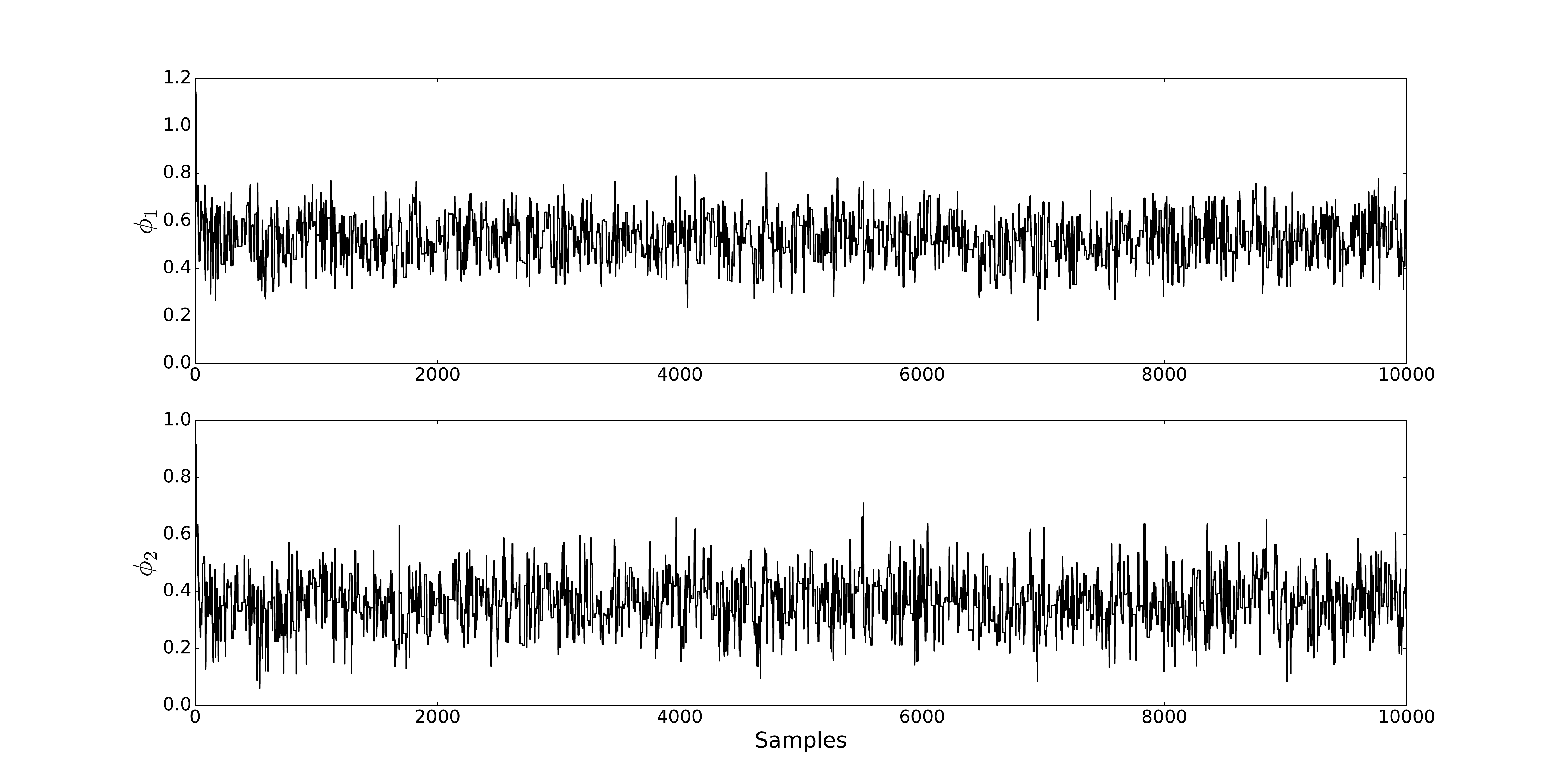}
	\caption{Markov chain for $\phi_1$ and $\phi_2$.}
	\label{fig:Markov_phi_2}
\end{figure}

\begin{figure}[h!]
	\centering
	\includegraphics[trim={3.0cm, 1.0cm, 3.0cm, 1.0cm}, clip, scale=0.3]{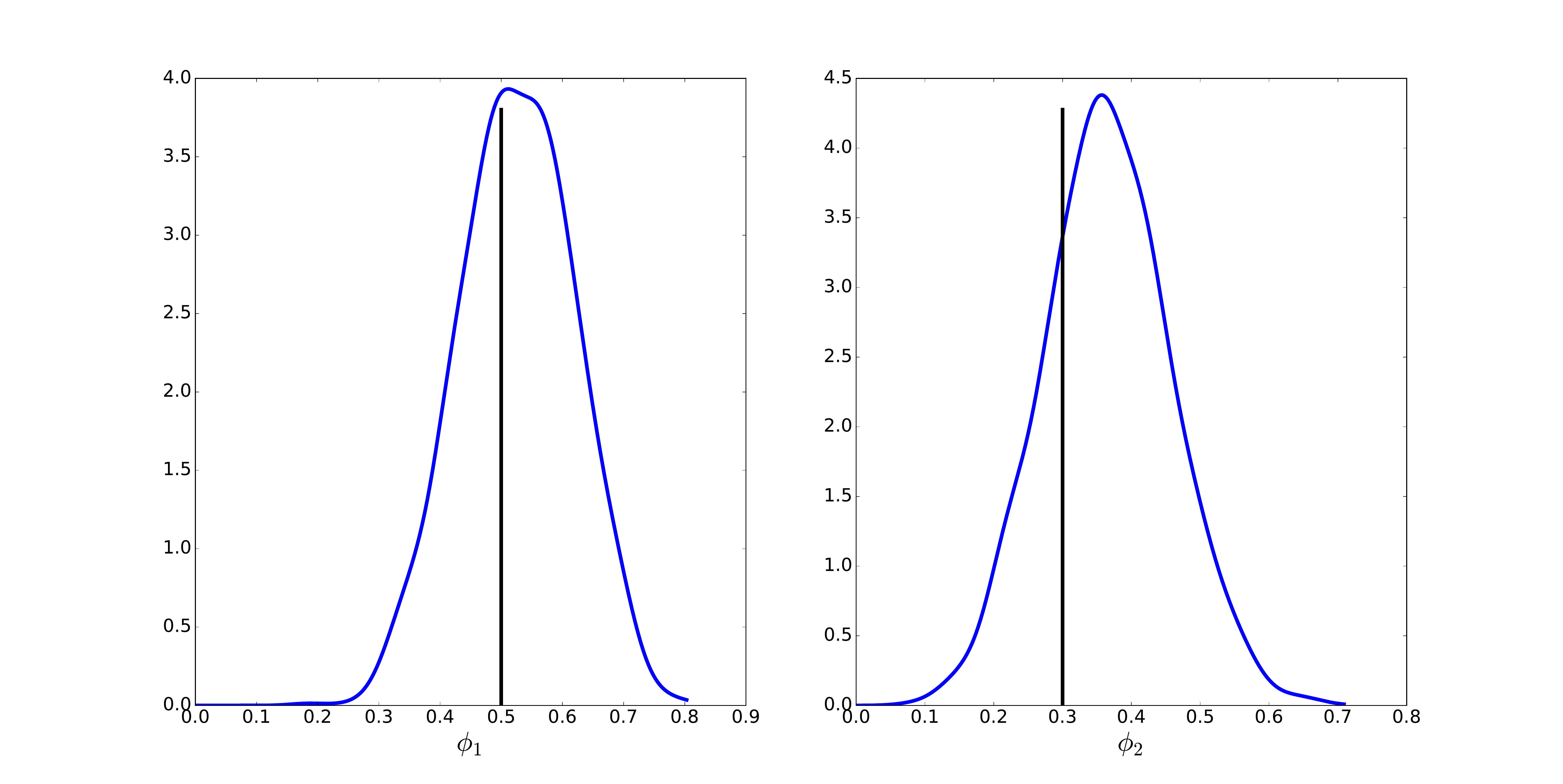}
	\caption{Kernel density estimation for the marginal posteriors of $\phi_1$ and $\phi_2$. Vertical bars show true simulated values.}
	\label{fig:kernel_density_2}
\end{figure}

\begin{figure}[h!]
	\centering
	\includegraphics[trim={3.0cm, 1.0cm, 3.0cm, 1.0cm}, clip, scale=0.3]{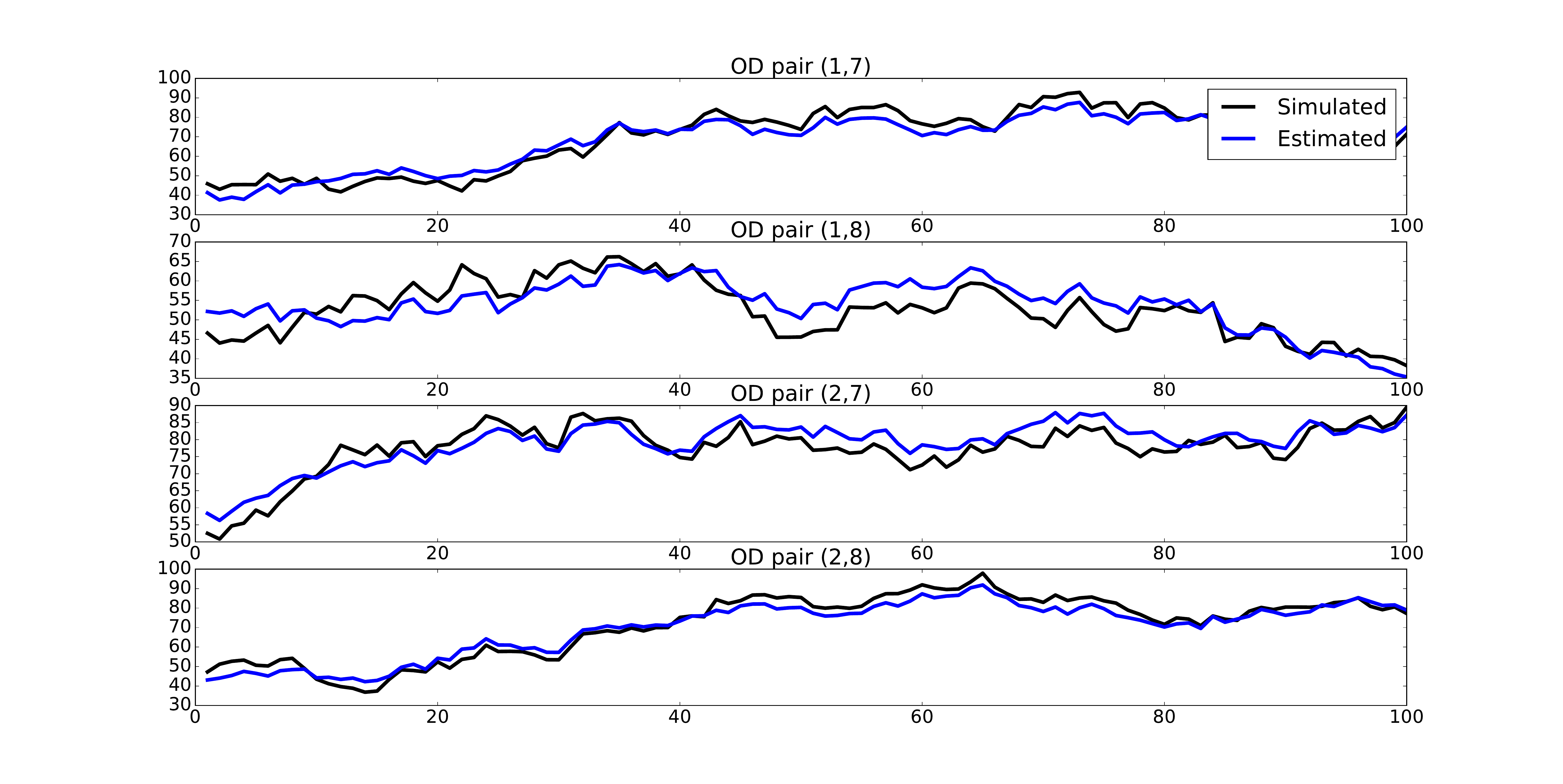}
	\caption{Simulated and estimated mean OD flows}
	\label{fig:odflows_2}
\end{figure}

The resulting Markov chain of sampled values is given in Figure \ref{fig:Markov_phi_2}. Figure \ref{fig:kernel_density_2} shows kernel density estimation of the marginal posteriors of $\phi_1$ and $\phi_2$ obtained from the sampled values. It can be seen that the sampled marginal posteriors have high densities around the true simulated values for $\phi_1$ and $\phi_2$. The sample mean values for $\phi_1$ and $\phi_2$ were 0.5250 and 0.3651, respectively, and highest posterior density regions with 95\% probability were [0.3546, 0.7067] and [0.1979, 0.5437], respectively. Figure \ref{fig:odflows_2} shows the simulated and estimated mean OD flows, with mean squared error (MSE) between simulated and estimated mean OD flows of 15.83. From these results, we can conclude that the proposed MCMC algorithm was able to estimate with good accuracy both mean OD flows and route choice parameters.

\subsection{Unknown evolution matrix}

Although in theory the evolution matrix $\matr{W}_t$ in DLMs may be jointly estimated along with other parameters, it is often treated as a known parameter provided by the analyst. A convenient way of specifying the evolution matrix is through discount factors \cite{west}. According to equation \eqref{eq:prior_mean}, the prior covariance matrix of the OD flows at time $t$ is given by $\matr{\bar{C}}_t = \matr{C}_{t-1}+\matr{W}_t$, i.e., it is the posterior covariance matrix at the previous time period amplified by the evolution covariance matrix, which corresponds to the increase in uncertainty due to time. Then, we can write the prior covariance matrix as $\matr{\bar{C}}_t = 1/\delta \matr{C}_{t-1}$, where $0 < \delta \leq 1$, so as to $\matr{W}_t = (1-\delta)/\delta \matr{C}_{t-1}$. The term $\delta$ is a discount factor. Notice that, when $\delta = 1.0$, we do not have an increase in uncertainty from time $t-1$ to $t$, thus corresponding to a static model. Lower $\delta$ values correspond to higher increase in uncertainty from time $t-1$ to time $t$.   

We ran our proposed MCMC algorithm with varying discount factors $\delta$, starting from covariance matrix $\matr{C}_0 = 1000 \, \matr{I}$, in order to assess its impact on the quality of the estimation of the route choice parameters and mean OD flows. In Table \ref{tab_discount} we show the corresponding sample means $\hat{\phi}_1$ and $\hat{\phi}_2$, 95\% highest posterior density regions for $\phi_1$ and $\phi_2$, and MSE between simulated and estimated mean OD flows.

\begin{table}[]
	\centering
	\caption{Estimation results given different discount factors (HPD - 95\% highest posterior density region)}
	\label{tab_discount}
	\begin{tabular}{cccccc}
		\hline
		$\delta$ & $\hat{\phi}_1$ & HPD & $\hat{\phi}_2$ & HPD & MSE(OD flows) \\ \hline
		0.7 & 0.4941 & [0.3265, 0.6659] & 0.3340 & [0.1475, 0.5007]  & 137.27   \\
		0.8 & 0.4960 & [0.3178, 0.6600] & 0.3380 & [0.1674, 0.5288]  & 82.84 \\
		0.9 & 0.4898 & [0.3121, 0.6606] & 0.3313 & [0.1480, 0.5057]  & 33.07 \\ \hline
	\end{tabular}
\end{table}

\begin{figure}[h!]
	\centering
	\includegraphics[trim={3.0cm, 1.0cm, 3.0cm, 1.0cm}, clip, scale=0.3]{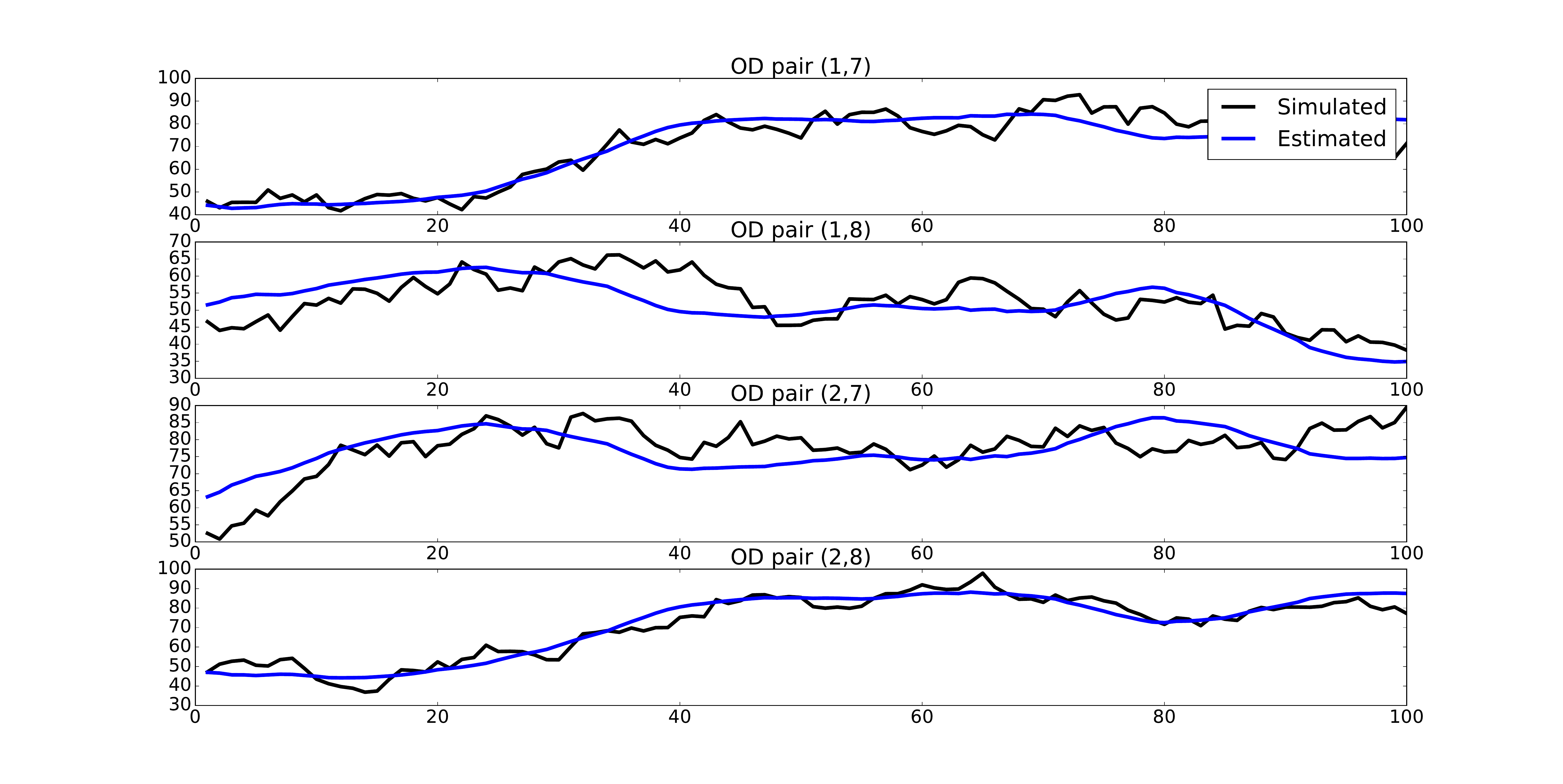}
	\caption{Simulated and estimated mean OD flows for $\delta = 0.9$}
	\label{fig:odflows_delta_09}
\end{figure}

Figure \ref{fig:odflows_delta_09} illustrates the effect of using $\delta = 0.9$ on the estimation of mean OD flows. As can be seen from the results, there was little difference in estimation quality between the cases with known and unknown evolution matrix regarding the parameters $\bm{\phi}$. The difference is more pronounced with relation to estimation quality of mean OD flows, which exhibited higher MSE with the evolution matrix specified by means of a discount factor.

\subsection{Observation of traffic volumes on partial links}

In some real transportation networks, there are data on traffic volumes only on a few links. We consider the application of our MCMC algorithm in these settings. In these set of experiments, all parameters are equal to the simulated ones, except for the mean OD $\bm{\theta}_{1:T}$ flows and the parameters $\bm{\phi}$ of the utility model, which are estimated by the MCMC algorithm. We vary only the number of links and in which of them we have traffic volume data.

We test three cases: observations on 1 link, on 2 links, and on 3 links in the network in Figure \ref{fig:net_hazelton_2015}. Links were selected so that we have a representative subset of the network. In Table \ref{tab:partial_links} we show the corresponding sample means $\hat{\phi}_1$ and $\hat{\phi}_2$, 95\% highest posterior density regions for $\phi_1$ and $\phi_2$, and MSE between simulated and estimated mean OD flows.

\begin{table}[h]
	\centering
	\caption{Estimation results with observation on partial links (HPD - 95\% highest posterior density region)}
	\label{tab:partial_links}
	\begin{tabular}{cccccc}
		\hline
		Observed links & $\hat{\phi}_1$ & HPD & $\hat{\phi}_2$ & HPD & MSE(OD flows) \\ \hline
		1 & 0.6088 & [0.0065, 1.1342] & 0.3904 & [0.0014, 0.8312] & 471.67 \\
		2 & 0.6283 & [0.3897, 0.8247]  & 0.4707 & [0.2509, 0.6827]   & 256.53 \\
		9 & 0.5877 & [0.3697, 0.7972] & 0.3644 & [0.1608, 0.5752] & 233.34   \\
		2 and 5 & 0.5115 & [0.3194, 0.6817]  & 0.3681 & [0.1692,  0.5422] & 246.70 \\
		1 and 9 & 0.5728 & [0.3608, 0.7456] & 0.3640 & [0.1760, 0.5546] & 175.43 \\
		2, 5 and 9 & 0.5023 & [0.3241, 0.6838]  & 0.3323 & [0.1383,  0.5039] & 86.51 \\		
		1, 7 and 9 & 0.5933 & [0.3947, 0.7936] & 0.3676 & [0.1870,   0.5870] & 59.82 \\ \hline
	\end{tabular}
\end{table}

\begin{figure}[h!]
	\centering
	\includegraphics[trim={3.0cm, 1.0cm, 3.0cm, 1.0cm}, clip, scale=0.3]{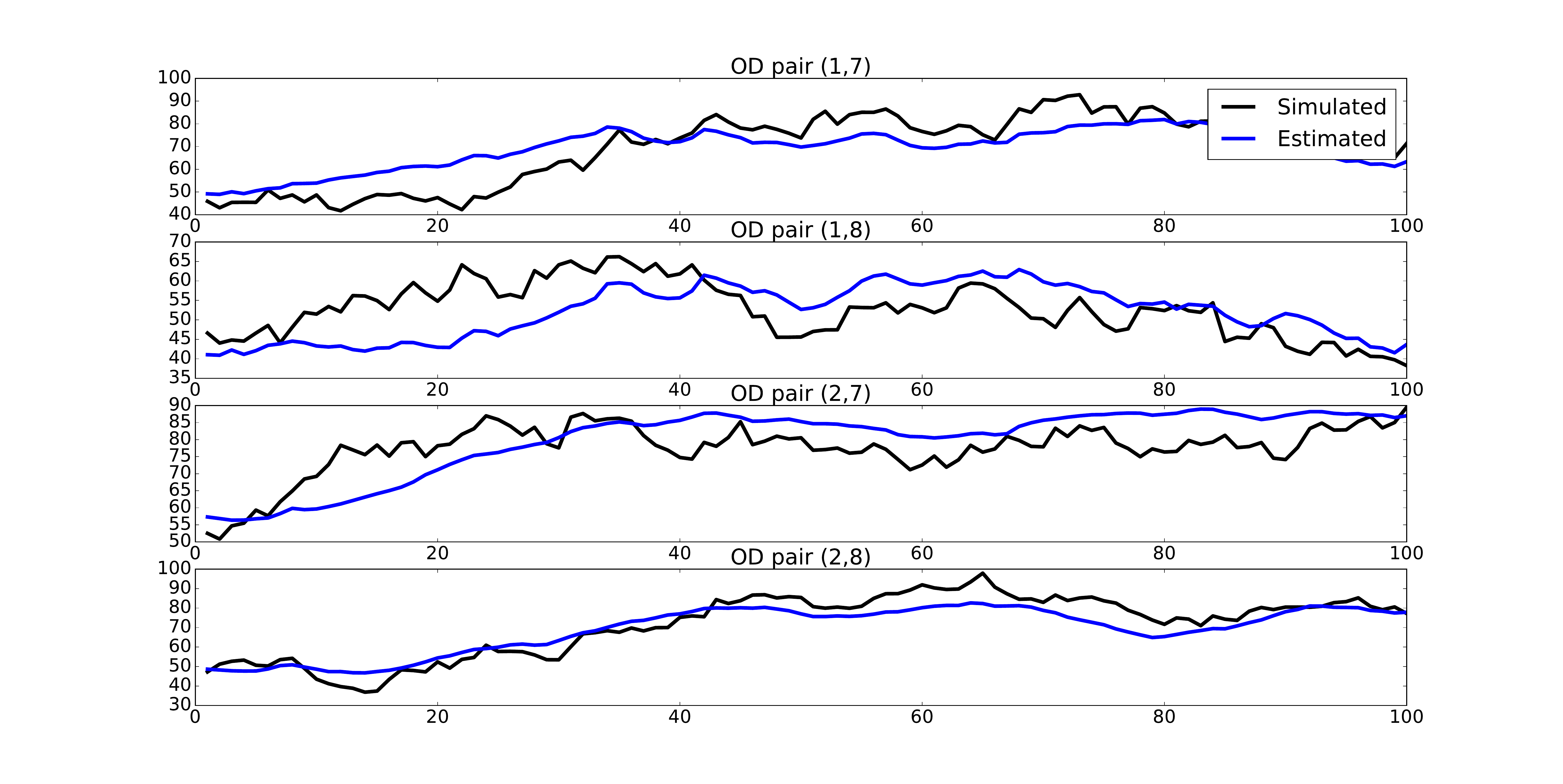}
	\caption{Simulated and estimated mean OD flows with observed traffic volumes on link 1, 7 and 9}
	\label{fig:odflows_link_1_7_9}
\end{figure}

In Table \ref{tab:partial_links}, we see that MSE of mean OD flows decreases as we observe more links. This result is in agreement with similar results of experiments with static models in the literature. The best results were obtained with data on links 1, 7 and 9, for which $\mathrm{MSE} = 59.82$. It is noteworthy that this result is not very far from the case studied in section \ref{sec:application}, where we have traffic volume data on all links and $\mathrm{MSE} = 15.83$. Figure \ref{fig:odflows_link_1_7_9} shows simulated and estimated mean OD flows for this case with data on links 1, 7 and 9.

Estimation quality of route choice parameters $\bm{\phi}$ also increases with the observation of more links, as the HPD regions get tighter around the simulated values $\bm{\phi} = (0.5, 0.3)^{\transp}$. Nevertheless, this effect is not so pronounced for OD flows. The best result among the tested cases was obtained when observing links 2, 5 and 9, which are links located in the central and more congested region of the network. It is also worth noting that we could estimate route choice parameters to a good accuracy by observing only link 9.

\begin{figure}[h!]
	\centering
	\includegraphics[trim={3.0cm, 1.0cm, 3.0cm, 1.0cm}, clip, scale=0.3]{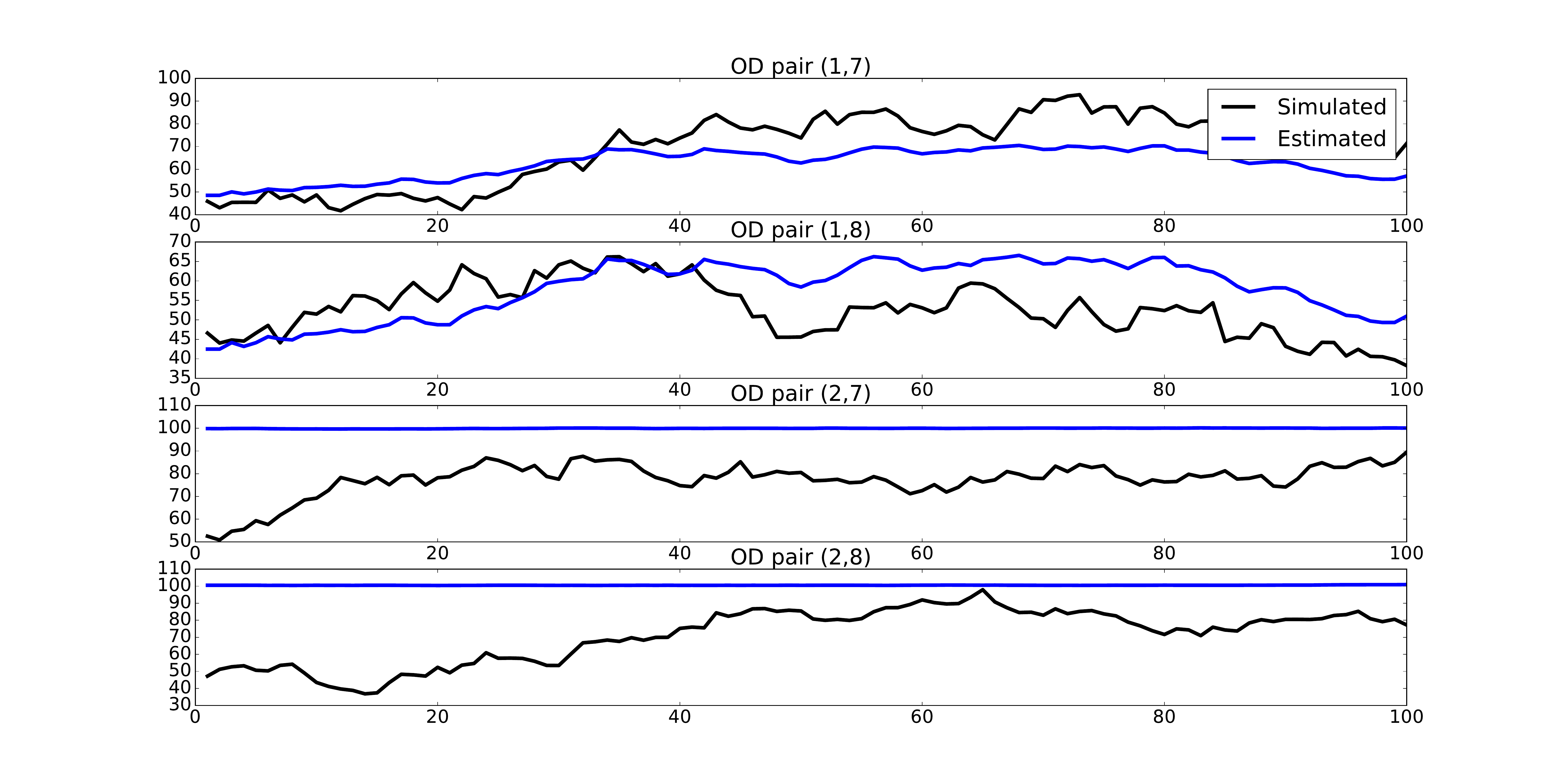}
	\caption{Simulated and estimated mean OD flows with observed traffic volumes on link 1}
	\label{fig:odflows_link_1}
\end{figure}

In addition, we notice that the estimation results for the case when we have traffic volumes data only on link 1 are the worst. This may be possibly due to the fact that there is no route in OD pairs (2,7) and (2,8) which includes link 1. Thus, traffic volumes on link 1 give no information on OD flows for these OD pairs. In Figure \ref{fig:odflows_link_1} we see that estimated mean OD flows remain at 100 in this case, which is our prior mean at $t=0$.

Finally, we should emphasize that we obtained good estimates when observing only a subset of links even though we used uninformative priors for both mean OD flows and route choice parameters. Static models often resort to prior knowledge, in the form of prior OD matrices, in order to ``regularize'' estimates when data are available only on a few links.

\section{Conclusion}
\label{sec:conclusion}

In this paper, we proposed a dynamic linear model for day-to-day OD flows in transportation networks. In our model, we took into account variability in OD flows, route choices and traffic volume measurements. We also modeled route choices through a utility model based on past route costs. We proposed a Markov chain Monte Carlo algorithm in order to sample from the joint posterior distribution of mean OD flows and route choice parameters by using data on traffic volumes on links and past route costs. Our model can be applied to congested networks and when data on only a subset of links are available.

We illustrated the application of the DLM and MCMC algorithm on a test network from the literature using simulated data. In our experiments, we were able to estimate with good accuracy both mean OD flows and route choice parameters using uninformative prior distributions and data on a subset of links in the test network. These are very promising results and indicate that dynamic linear modeling of day-to-day OD flows may provide a valuable tool in analyzing and estimating OD demand in transportation networks.

As further extensions to our work we suggest the consideration of trend and seasonality in OD flows. This can be implemented through the definition of additional parameters such as trend and seasonal factors, whose relationship with mean OD flows can be easily established by means of the system matrix $\matr{G}_t$. Another promising research direction is the use of our proposed DLM to forecast future traffic volumes, after setting up the model with estimated route choice parameters.

\pagebreak

\bibliographystyle{abbrv}
\bibliography{References}
\end{document}